# Radiation detection and energy conversion in nuclear reactor environments by hybrid photovoltaic perovskites

Gábor Náfrádi[b,c,1], Endre Horváth[a], Márton Kollár[a], András Horváth[b], Pavao Andričević[a], Andrzej Sienkiewicz[a,d], László Forró[a], and Bálint Náfrádi[a]

[a]*Laboratory of Physics of Complex Matter, Ecole Polytechnique Fédérale de Lausanne (EPFL), CH-1015 Lausanne, Switzerland*

[b]*Institute of Nuclear Techniques (NTI), Budapest University of Technology and Economics (BME), H-1111 Budapest, Hungary*

[c]*ISIS Facility, Rutherford Appleton Laboratory, Chilton, Didcot, Oxfordshire OX11 0QX, United Kingdom*

[d]*ADSresonances Sàrl, CH-1028, Préverenges, Switzerland*

**Abstract**

**Detection and direct power conversion of high energy and high intensity ionizing radiation could be a key element in next generation nuclear reactor safety systems and space-born devices. For example, the Fukushima catastrophe in 2011 could have been largely prevented if 1% of the reactor's remnant radiation (γ-rays of the nuclear fission) were directly converted within the reactor to electricity to power the water cooling circuit. It is reported here that the hybrid halide perovskite methylammonium lead triiodide could perfectly play the role of a converter. Single crystals were irradiated by a typical shut-down γ-spectrum of a nuclear reactor with $7.61 \times 10^{14}$ Bq activity exhibit a high-efficiency of γ-ray to free charge carrier conversion with radiation hardening. The power density of 0.3 mW/kg of methylammonium lead triiodide at 50 Sv/h means a four times higher efficiency than that for silicon-based cells. The material was stable to the limits of the experiment without changing its performance up to 100 Sv/h dose rate and 57 Sv H*(10) ambient total γ-dose. Moreover, the γ-shielding performance of methylammonium lead triiodide was found to be superior to both ordinary and barite concrete.**

## 1. Introduction

Organic-inorganic Hybrid Halide Perovskites (HHPs) resulted in recent significant breakthroughs in a variety of optoelectronic applications. The flagship application has been a highly efficient photovoltaic cell based on the remarkable photo-absorbing and photo-conducting material - methylammonium lead triiodide, ($CH_3NH_3PbI_3$ here after, $MAPbI_3$). In particular, the power conversion efficiency of $MAPbI_3$-based solar cells has witnessed a stunning increase from the initial 3.8%[1] to over 25% nowadays[2]. HHP materials have also been foreseen for applications in lasers because of the low lasing threshold[3] and wavelength tunability[4], in light emitting diodes[5] and photodetectors[6] working in the entire visible spectral range[7]. The outstanding chemical tunability of the HHPs resulting from the possibility of replacement of I atoms by Br or Cl atoms in the initial $MAPbI_3$ compound allowed for the band gap engineering with enhanced optical response in the entire visible[8] and near-infrared ranges[9]. Self assembly-based micro fabrication[10] allowed nanometer-sized[11] and high-sensitivity[12] detector development. Recently, even magnetic compounds have been synthesized based on MAPbI$_3$ by substituting Pb atoms by magnetic atoms such as Mn[13]. These magnetic compounds are potential candidates for novel magneto-optical data storage and memory operations[13].

---

[1] corresponding author, gbrnfrdi@yahoo.com, present address is c.

It is also noteworthy that due to the presence of a heavy element, Pb, and heavy-halide content, the HHP materials exhibit an excellent absorption of X-rays[14] and γ-radiation[15]. Furthermore, calculations predicted that other forms of ionizing radiation, like neutron or electron, important amongst other examples in space applications can be detected and even used for energy harvesting[14]. A typical application could be in nuclear batteries beside nuclear power plants. Nuclear batteries use several methods, such as thermal conversion in radioisotope thermal generators (RTGs), or indirect conversion where two-step conversion happens. In indirect conversion, first usually a charged particle is converted to light (e.g. ultra violet (UV) radiation) which is later collected with photodiodes[16]. A third group of nuclear batteries uses various materials for direct harvesting of radiation, like alfa voltaic or beta-voltaic cells for example based on aluminium indium phosphate (Al$_{0.52}$In$_{0.48}$P)[17]. This type of device was tested with 182 MBq Fe-55 X-ray emitter (5.8 keV) and 185 MBq Ni-63 beta emitter (66.9 keV) radioisotopes. Despite the low energy irradiation, the reported conversion efficiencies were 2.2% and 0.06% respectively[17]. Beta-voltaic devices were reported with even higher efficiencies such as a silicon carbide (SiC) p-i-n junction based device had a 4.5% conversion efficiency upon to P-33 beta irradiation (248 keV)[18]. Possibly the oldest nuclear battery technology is the direct charge batteries[19] where the charged alfa or beta particles directly drive the current. This technology is present in Self Powered Neutron Detectors (SPNDs) as well.

In small size nuclear batteries or micro batteries, the high gamma energy emitter source is not favorable since the radiation shielding would be a problem. This means the energy per particle or photon is limited.

The initial success in device development is shadowed with parallel reports of chemical,[20] thermal[21] and temporal[22] instability of the devices together with the engineering difficulties of heat conduction[23], extraction[24] and mechanical brittleness of the structure[20]. In fact, in highly energetic environment, sever degradation is expected due to radiation damage[22]. Yet, experimental reports on the impact of radiation are largely controversial in the perovskite community. For example, low-dose and low dose rate soft X-rays have been reported to degrade MAPbI$_3$ to lead iodide (PbI$_2$)[1], while other reports indicated remarkable stability in the same energy range even for an extended radiation dose or dose rate[25]. There are also other reports pointing to an enhancement of not only the stability[26] but also the performance due to the exposure of the device to ionizing radiation[27].

MAPbI$_3$ already demonstrated good response for Cs-137 gamma radiation (661 keV)[15], therefore a further step was to irradiate the HHP samples in a pure gamma field in a shut down fission reactor spectrum where the highest photon energies are in the few MeV ranges and the dose rates are also higher. In the reactor core the remnant radiation of the uranium fuel produces α and β particles and γ radiation. The α, and β particles are efficiently absorbed by the fuel itself, whereas the γ photons, originating from the relaxation of the excited states of fission products, are the major concerns, since they are able to leave the reactor core.

Excellent gamma radiation harvesting potential of the HHP crystals is measured in an intense remnant gamma radiation field of a nuclear reactor. The irradiated crystals in the measurement did not show any sign of degradation.

The paper also highlights the unique possible triple function (detection, harvesting, shielding) of the materials showing an example of usage in passive safety systems in high gamma dose rate fields around nuclear reactors.

## 2. Methods

This work focuses on the gammavoltaic response of HHPs, that are MAPbI$_3$, methylammonia lead tribromide (CH$_3$NH$_3$PbBr$_3$ hereafter MAPbBr$_3$) and methylammonia lead trichloride (CH$_3$NH$_3$PbCl$_3$ hereafter MAPbCl$_3$) single crystal based devices in a highly energetic nuclear radiation environment in the vicinity of the core of a nuclear reactor just after a shut-down.

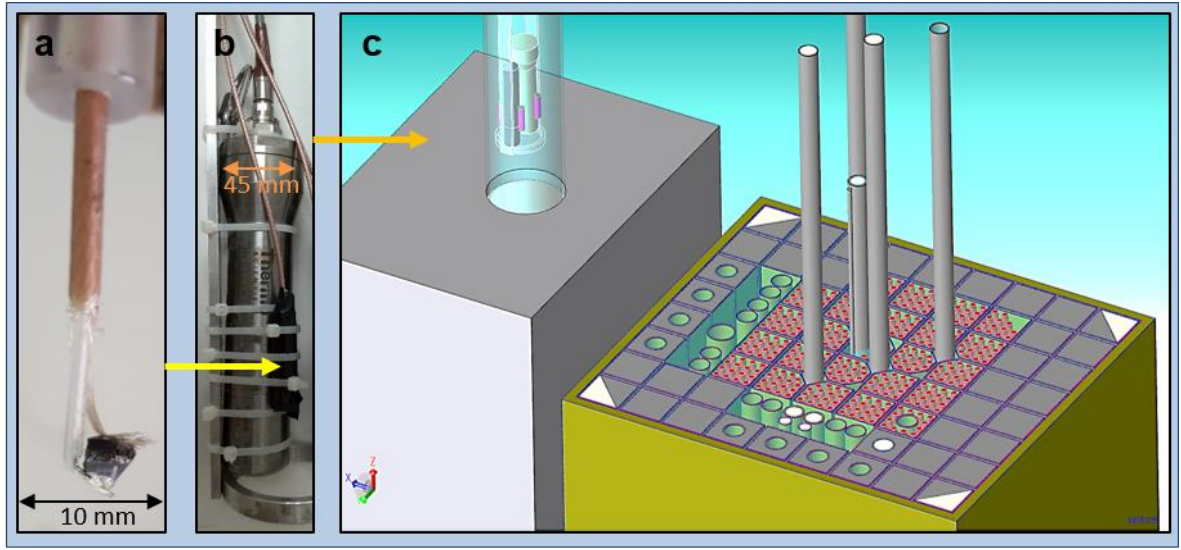

*Figure 1: Experimental configuration for the γ-exposure of HHPs in a nuclear reactor. **a**) A MAPbI$_3$ single crystal attached with conducting silver paste to Cu electrodes. **b**) The hermetically encapsulated and blindfolded samples (to prevent humidity induced degradation and potential artefacts from visible light exposure) are fixed to the γ radiation detector (indicated by the yellow arrow) prior to inserting the setup into the irradiation channel in the vicinity of the reactor core (orange arrow). **c**) Schematic view of the reactor core model with the dry irradiation channel, and the probe head in the irradiation position. The reactor core is inside the yellow area, it contains the fuel rods (the U-235 fission elements in red) organised into fuel assemblies, graphite reflectors (grey region around the fuel assemblies), various irradiation channels (green tubes) and control, safety rods and pneumatic rabbit system (grey long tubes in the middle of the core).*

The gammavoltaic measurements were carried out in the Training Reactor (TR) of the Budapest University of Technology and Economics (BME) in the shutdown state, just after a 1.35 h long 100 kW operation (Figure 1). MAPbBr$_3$, MAPbCl$_3$ and MAPbI$_3$ samples were irradiated simultaneously. This type of operation produces a pure gamma radiation field with $7.61\times10^{14}$ Bq initial radioactivity around the core which significantly exceeds the $10^5$-$10^7$ Bq range usually used for similar tests (See Table 1 in Appendix for an overview). The time and spectral evolution of the radiation field is similar to the radiation field in a Nuclear Power Plant (NPP) after the reactor shutdown.

Details of the single crystal growth, the irradiation and the gammavoltaic measurement can be found in sections 2.1. and 2.2. The simulations of the gamma spectrum evolution as a function of reactor cooling time can be found in section 2.3. The description of Monte Carlo N-Particle (MCNP)[28] based shielding calculations can be found in section 2.4. From the measured photocurrent and gamma dose rate the estimated photovoltaic power density extraction capacity of MAPbI$_3$ was calculated, the details can be found in section 2.5. Based on the power density, an estimation of potential power harvesting was given for large scale NPP -like environments, the details can be found in section 2.6. The stability of the samples was checked with photo current measurements, X-ray powder diffraction and photoluminescence measurements. Details can be found about these techniques at sections 2.7-2.9.

### 2.1. Sample production and sample preparation

Crystals of $CH_3NH_3PbI_3$ were synthesized by solution growth. The 20.0 g lead(II) acetate trihydrate ($Pb(ac)_2 \times 3H_2O$, >99.9%) was reacted with 96.0 ml saturated hidrogen iodide (HI) solution (57 wt% HI in water ($H_2O$)). The formed $PbI_2$ precipitate was dissolved in the acidic solution by shaking. The respective amount (52.7 mmol) methylamine ($CH_3NH_2$) solution (40 wt% in $H_2O$) was pipetted into the 5 °C ice-cooled solution of $PbI_2$. The cold solution avoids evaporation of methylamine during the exothermic reaction. Dark coloured microcrystallites of $MAPbI_3$ were formed immediately and settled down at the bottom of the vessel. In a temperature gradient of 0.5 °C/cm in the acidic media, large (3–10 mm) black crystals were formed in 7-10 days. The $CH_3NH_3PbBr_3$ and the $CH_3NH_3PbCl_3$ were synthesized using the same method, but the stoichiometrically needed hydrogen bromide (HBr) solution (48 wt% HBr in $H_2O$) and hydrogen chloride (HCl) solution (36 wt% HCl in $H_2O$), respectively.

*2.2. Irradiation setup*

For γ-irradiation the TR BME was used. The reactor was shut-down after 1.35 h 100 kW operation. This type of operation produces a pure gamma radiation field around the core. The time and spectral evolution of the radiation field is similar to the radiation field in a NPP after shutdown, except that the structural components of TR are made of aluminium. To avoid the effect of γ photons from aluminium structural components the experiments started after about 20 mins of cooling of the core. This cooling time ensures that the Al-28 isotope will practically totally decays (the half life of Al-28 is 2.24 min).

For irradiation the samples and the dose rate meters were fixed on an irradiation probe in close vicinity to each other (Figure 1). In situ active dose rate measurement was performed by a Thermo Scientific FHZ-312 detector. Rotem AMP-200 dose rate meter was used as backup detector. The samples (Figure 1) were placed in a small transparent plastic holder, which was sealed to prevent degradation due to air and humidity. The $MAPbI_3$ sample size was about 0.034 $cm^3$, the characteristic edge length was 3.24 mm.

The samples were blind folded with black duct tapes to avoid response to visible light. During the irradiation the generated photocurrent of the $MAPbI_3$ and $MAPbBr_3$ crystals was measured with 10 V bias voltage while the $MAPbCl_3$ was measured with 5V bias. The current of the $MAPbI_3$ and $MAPbBr_3$ samples was measured by a Keithley 2400 source meter. For $MAPbCl_3$ a Kethley 2100 was used. Hameg HM8040 triple power supply was used together with the Kethley 2100. While the Keithley 2400 source meter was used as a source (power supply) and a meter simultaneously. The irradiation probe was inserted into a vertical dry irradiation channel of the reactor (Figure 1). The probe was lowered until the measured dose rate of the detector reached its limit: 100 Sv/h. The probe was then fixed in a higher position, 72.1 cm above the bottom of the irradiation channel, to keep the dose rate meters under their detection limit. The whole set up was kept in this position until the end of the measurement.

*2.3. Neutronics calculations*

Neutron and gamma transport calculations were also performed with MCNP code to determine the gamma spectrum which was present during the irradiation and to determine the time evolution of the gamma spectrum. The MCNP model of the reactor core was developed by using SuperMc[29] which incorporates MCAM software[30], the interface program for CAD to MCNP input conversion. From the reactor power history, first the isotope composition of the fuel was calculated by using the BURN card of MCNP. From the isotope composition the source gamma lines and intensities were obtained using a code called Microshield[31]. Only the isotopes with greater than 1 kBq were taken into account, this simplifications still resulted in more than 50 different isotopes. Microshield generated the source spectrum with 19 discrete energy lines between 15 keV and 3 MeV. A second MCNP calculation used the gamma source spectrum obtained from Microshield and calculated the evolved photon spectrum in the irradiation position. The calculations were repeated at five different cooling times: 0 min, 15 min 1 hour, 2 hours and 3 hours after the shutdown. The spectrum does not contain the gamma line radiation from the aluminium structural components. The obtained photon spectra can be seen in Figure 2b. The spectrum is varying in time as the different isotope chains are decaying. The maximum of the spectrum is located around small energies, less than 700 keV, where the potential absorption of small MAPbI$_3$ samples is maximal[14].

*2.4. Shielding comparison calculations*

The shielding comparison calculations done with MCNP were carried out with a 10 cm thick cylindrical shielding cell, with a radius of 50 cm. The initial photons hit the shielding cell perpendicularly, and the cylinder surface was set to reflective. For photons the mcplib84 data table was used. For electrons, the el03 data table was used. The photon flux to dose rate conversion was done by applying the ICRP-21 guidelines[32]. Material compositions were extracted from Compendium of Material Composition Data for Radiation Transport Modeling[33]. The MAPbI$_3$ density in the MCNP calculations was 4.1082 g/cm$^3$.

*2.5. Power density extraction evaluation*

The photo current was measured with 10 V bias, with 1 V only 10% of it would be measured. The power density can be calculated from the photo current. At $D$=48.9 Sv/h dose rate in the $m$=0.14 g sample the radiation would generate $I_{PH}$=42.054 nA photocurrent with $V$ = 1 V bias potential. The extracted power density is $P_e$=$I_{PH}$*$V$/$m$= 0.3 mW/kg. The efficiency is the ratio of the output power density and the deposited power density. The deposited power density (or dose rate) was 13.58 mW/kg therefore the efficiency is 0.3 mW/kg/13.58 mW/kg=0.022=2.2%.

*2.6. Nuclear Power Plant power extraction estimation*

MSv/h dose rate radiation field can be present, for example inside the active core of a nuclear reactor. The radiation intensity is decreasing by 1/$d^2$ for a point source and 1/$d$ for line sources ($d$ is the distance from the source). It allows harvesting the same emitted energy at a small radius with high dose rate or at a larger radius with lower dose rate by applying a larger area harvesting system. Assuming 4500 kGy/h gamma dose rate in a core of a reactor and 1500 kGy/h at 50 cm distance from it. The distance can be found where the dose rate will be about 50 Gy/h which enables to directly use the measured response to calculate the energy harvesting potential. In a non-absorbing medium this distance would be 86.6 m for 1/$d^2$ and 15 km for 1/$d$. This distance is much larger than the actual size of the reactors due to the applied absorbing shielding. Usually the 50 Sv/h is reached inside the shielding around the core at an axial distance of 0.5 m - 1 m from the edge of the core. (Figure 5.2 of[34] ). Without shielding the dose rate at 1 m from the edge of the core would be about 375 kGy/h considering the 1/$d^2$ decrease.

A more realistic estimation could be the following: cover a reactor core with shielding, followed by an additional thick layer of MAPbI$_3$. Considering a Canada Deuterium Uranium (CANDU) 6 reactor with a core diameter of 7.6 m and a core length of 6 m and assuming 1 m of concrete shielding before reaching the 50 Sv/h dose rate in every direction from the core it will end up in a cylinder surface of 386.03 m$^2$. Applying 10 cm of MAPbI$_3$ layer after the shielding as a harvester it requires 38.60 m$^3$. Using the 4.2864 g/cm$^3$ density one can obtain 165471.7 kg of MAPbI$_3$. Using the 0.3 mW/kg power density the output power of such a system would be 49.64 W. So after 1 m of absorbing shielding layer about 0.05 kW could be still harvested with such a system. On the other hand, a significant amount of radiation and power (99.98% or 374.95 kGy/h, considering the calculations above) is still wasted in the shielding before the perovskite layer. Utilizing that portion even with less efficiency could end up in kilowatts.

*2.7. Photo-current measurements*

Photocurrent measurement was performed at room temperature before and after irradiation of the HHP single crystals. The crystals were contacted symmetrically with silver epoxy and copper wires. The current was measured by a Keithley 2400 source meter. The dark current of MAPbI$_3$ stabilized asymptotically at about 200 nA at 10 V bias and it was independent of the radiation history. The samples reacted to visible light radiation, with 10 V bias then the samples were blind folded with black duct tape covering. Four measurements were performed, a pre-irradiation on-off cycling to obtain the photo-response, a prior irradiation reference dark-current measurement without irradiation, a current measurement during a high gamma dose rate irradiation and an on-off cycling after the irradiation. The dark current measurement prior to irradiation and the irradiation measurements were carried out with blind folded samples. Photocurrent experiments before irradiation were performed at 1 V bias. After the irradiation experiments were performed at both 1 V and 10 V bias. Due to the ohmic character of the symmetrically contacted devices the photo-response is linear, thus the 10 V equivalent photocurrent is reported through the manuscript.

*2.8. X-ray powder diffraction*

The MAPbI$_3$ crystallite was powdered after the irradiation in order to check if additional phases, namely if PbI$_2$, ( product of degradation) appeared in it. The powder X-Ray Diffraction (XRD) patterns of the samples were obtained using a PANalytical Empyrean XRD with Bragg-Brentano geometry fitted with a PANalytical PIXcel-1D detector using Ni filtered Cu-K$\alpha$ radiation ($\lambda$ = 1.54056 Å).

*2.9. Photoluminescence*

Phtoluminescence spectra were recorded using a customized inverted biological epi-fluorescent microscope (TC5500, Meiji Techno, Japan), which was combined with a commercially available spectrofluorometer (USB 2000+XR, Ocean Optics Inc., USA). This setup enabled to simultaneously acquire photoluminescence microscopic images and the corresponding spectra of MAPbX$_3$ single crystals (X denotes Cl, Br or I) under excitation with the monochromatic incoherent light at 405 nm, 470 nm and 546 nm wavelengths.

## 3. Results and discussion

The maximum dose rate in the experiments was over 100 Sv/h, which is beyond the saturation level of the gamma detectors, thus the data collection started once the reactor core cooled below ~50 Sv/h dose rate. The measured γ dose rate and γ-induced photo-current as a function of time are plotted in Figure 2. The fit of the γ-dose rate gives an exponential decay, as expected in the shut-down state, with a time constant of $\tau^\gamma$ = 3750 ± 20 s, (green line in Figure 2). The trend of the measured photocurrent decay followed that of the γ dose-rate, however, with a slightly reduced time constant $\tau^{PC}$ = 2290 ± 13 s (Figure 2a). This is the best seen in the upward curvature of the γ-induced photocurrent normalized by the γ-dose rate shown in the inset. The reason for the decreased $\tau^{PC}$ relative to $\tau^\gamma$ could be read from Figure 2b where the gamma spectrum at different time-points is determined by MCNP calculations adopted to the experimental conditions. One can notice that the remnant radiation of the reactor produces an unsteady γ-spectrum after the reactor shutdown. The number of γ-photons versus their energy is plotted at 0, 15, 60, 120 and 180 min time-points. Both the γ-dose rate and the spectrum change as a function of time. The origin of this time variation is the presence of multitude of radioisotopes in a nuclear fission. In the nuclear reactor fuelled with U-235, in the sequence of fission, over 50 major radioisotopes with different half-times are created. Accordingly, due to the different half-times, the spectrum of the γ radiation field in parallel with the decreasing dose rate change with time (Figure 2b). During the experiment, the radiation field softens, there are relatively more photons at lower energies and at longer cooling times. The low-energy photons are more efficiently absorbed by the HHP-based device, which results in a higher photocurrent as shown in the inset of Figure 2a.

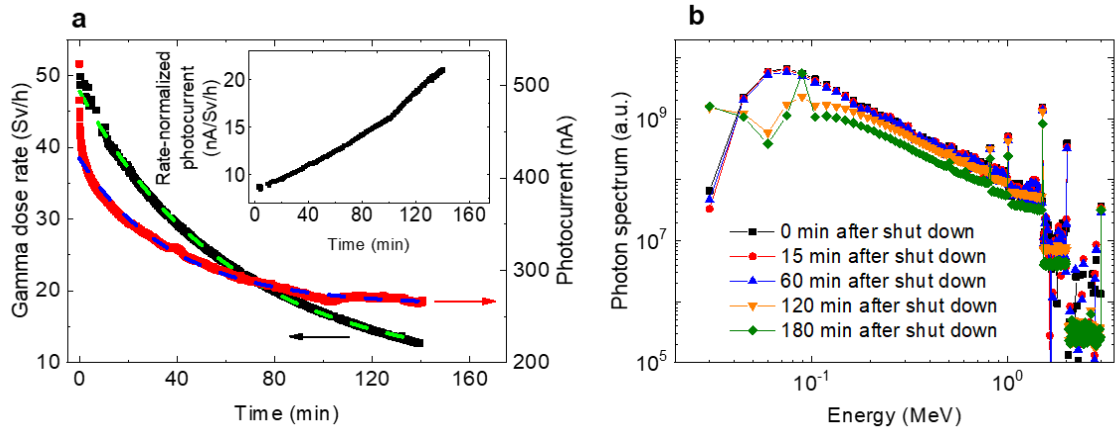

*Figure 2: Gammavoltaic effect of MAPbI$_3$ exposed to the γ radiation of a shutdown nuclear reactor a) Time dependence of the measured gamma dose rate in black (left axis) and the resulting photocurrent in red (right axis) across the MAPbI$_3$ single crystal placed into the core of a shutdown nuclear reactor. The γ dose rate decays exponentially as expected for the reactor core as cooling down; the fit (green dashed line) gives a time constant of $\tau^{\gamma}=3750\pm20$ s. The γ-induced photocurrent follows the decreasing γ-dose rate with $\tau^{PC}=2290\pm13$ s (extracted from the fit in blue dashed line). Inset shows the time dependence of the photocurrent normalized with the dose rate. b) Calculated γ spectrum by MCNP code at the sample position at several time points.*

One of the central findings of the study is that at a 48.9 Sv/h dose rate, a 0.3 mW/kg output power density can be reached by the 3.24 mm thick MAPbI$_3$ devices, with a 3 V/mm potential difference. The potential difference was provided by an external power supply. With asymmetrical contacts much larger built in fields could be generated thus this number is a conservative underestimate.[35] Nevertheless, 0.3 mW/kg is a rather significant energy harvesting potential. In the vicinity of the core of NPP-s up to a few million Sv/h dose rates can be present[34], which extrapolates to a few kW/kg energy production, assuming the conversion efficiency remains unchanged. If a 10 cm thick MAPbI$_3$ layer would be placed behind the shielding, in cylindrical surface around a nuclear reactor, where 50 Sv/h dose rate gamma field is present, the 0.3 mW/kg output power density could add up to about 0.05 kW of electric power.

Achieving 0.3 mW/kg output power density means a four times higher efficiency in respect to the state-of-the-art Si-based photovoltaic cells where 0.34 mW/kg was achieved at four times larger ~200 Gy/h energy deposition rate from a Co-60 source[1]. The estimated efficiency of the MAPbI$_3$ sample at 48.9 Sv/h dose rate is 2.2%. The improved performance is attributed to a combination of several factors. The optical band-gap of MAPbI$_3$ is about half of that of Si ($\Delta_\Gamma=3.4$ eV). This results a factor of two increase in photo-electron generation[36]. The active volume of a device is characterized by the 10$^{th}$-value layer (Figure 3) is also comparatively smaller for MAPbI$_3$ facilitating the effective charge collection. In addition, the charge extraction, even for large MAPbI$_3$ crystals, is rather large over 75% of all photoelectrons can be collected with a low (1-10 V) bias voltage.[14]

It is important to note that the response of MAPbI$_3$ for unit dose rate is larger at low dose rate (Figure 2a inset), but not saturating even at the highest dose rates. This indicates that even in larger dose rates MAPbI$_3$ devices could potentially be used.. This remarkable behaviour implies that MAPbI$_3$ is a good candidate for γ dose rate measurements or for direct gammavoltaic energy harvesting relevant at nuclear power plants or even at radioactive waste storage environment. To what extent of the few million Sv/h dose rates of shutdown NPP radiation can be used, however, requires further stability studies. Nevertheless, positioning the HHP-based elements to an appropriate distance from the NPP core can ensure long term device stability.

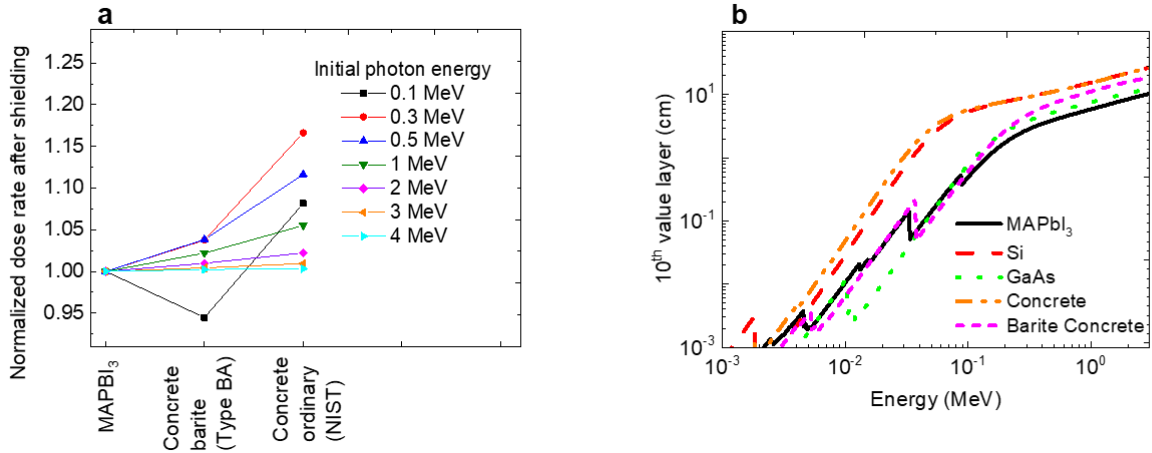

*Figure 3: Radiation shielding properties of MAPbI₃ a) Comparison of normalized dose rate shielded by 10 cm thick slab of MAPbI₃, BA type barite concrete and ordinary concrete calculated by MCNP. Different colours represent initial photon energy. b) Calculated 10$^{th}$ value layer, i.e. the layer where over 90% of the radiated energy is absorbed, for MAPbI₃, Si and GaAs (materials for photon-detection) and various concretes used as shielding materials.*

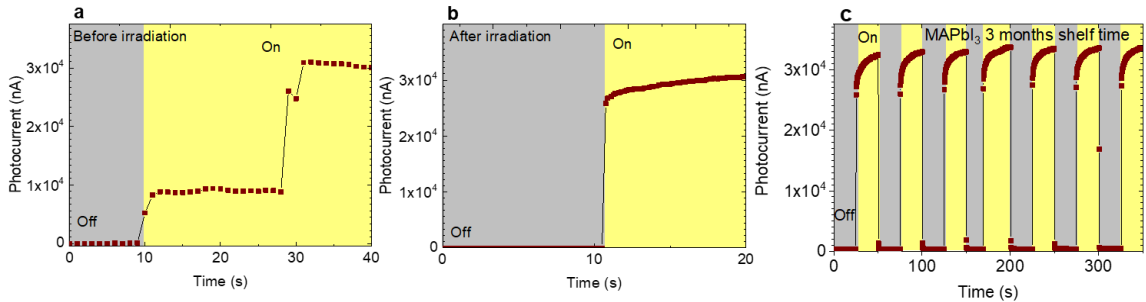

*Figure 4: Optical response of the MAPbI₃ devices before and after the γ exposure. a) Photocurrent measured to the MAPbI₃ device before irradiation exposed to a white light source. b) Photocurrent extracted from the same MAPbI₃ crystal after its exposure to the γ-spectrum of a nuclear reactor with 7.61×10$^{14}$ Bq activity. c) The photo-response of the device shows no degradation even after three months shelf time.*

Apart from the high efficiency energy harvesting from γ radiation, other important criteria for applications is the material stability at high dose rates and the radiation-hardened behaviour. To verify the radiation hardness of MAPbI₃ devices the optical response of the device was compared before and after the 57 Sv H*(10) ambient γ-dose irradiation. Note, that this dose which corresponds to 7.61×10$^{14}$ Bq initial activity is of about 20-times higher than the median human lethal radiation dose[37] and significantly exceeds the previously reported activity values (see Table. 1). Figure 4 shows that the photocurrent generated upon white light illumination before (Figure4a) and after γ irradiation (Figure4b) and after three months of a shelf (Figure 4C), within the precision of the measurement, is the same. Note the same photocurrent scale: the device shows no sign of degradation in the photo-response and the response speed seems also similar. This indicates that MAPbI₃ survived the 57 Sv H*(10) ambient γ-dose without any sign of degradation. This observation corroborates with powder X-ray diffraction measurement, which shows no traces of PbI₂ or other degradation products after irradiation (Figure 5). Furthermore, photoluminescence measurements also confirmed the absence of radiation damage (Figure 6). The relatively narrow ( Full Width at Half Maximum (FWHM) = 50 nm) and unstructured peak centred on 770 nm, characteristic for MAPbI₃, points to a low level of radiation-induced damage to the crystal structure. It should be noted that the remarkable radiation hardness of MAPbI$_3$ seems to be universal across the HHPs. Besides MAPbI$_3$ MAPbBr$_3$ and MAPbCl$_3$ based devices were irradiated, as well. No sign of degradation was found in the optical responses (Figure 7 and Figure 8).

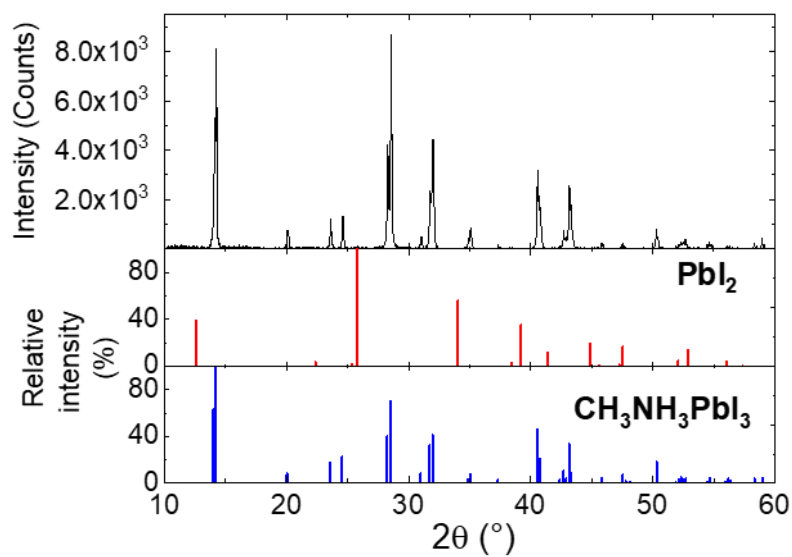

*Figure 5 – Powder X-ray diffractogram of MAPbI$_3$ after irradiation*

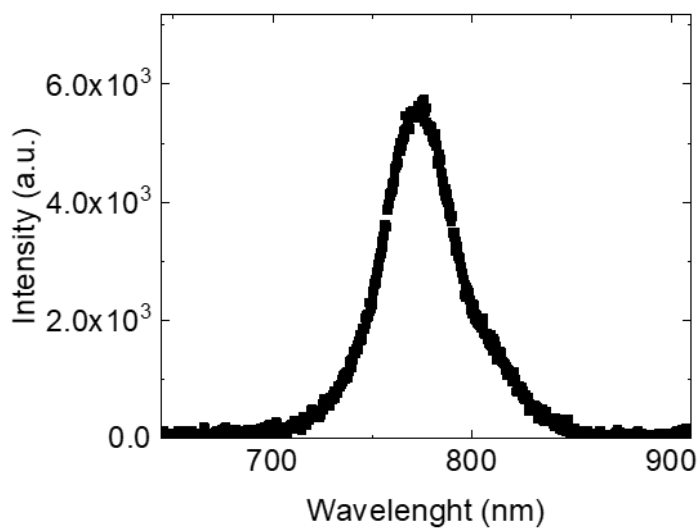

*Figure 6 – Photoluminescence spectrum of the MAPbI$_3$ after exposure to γ irradiation.*

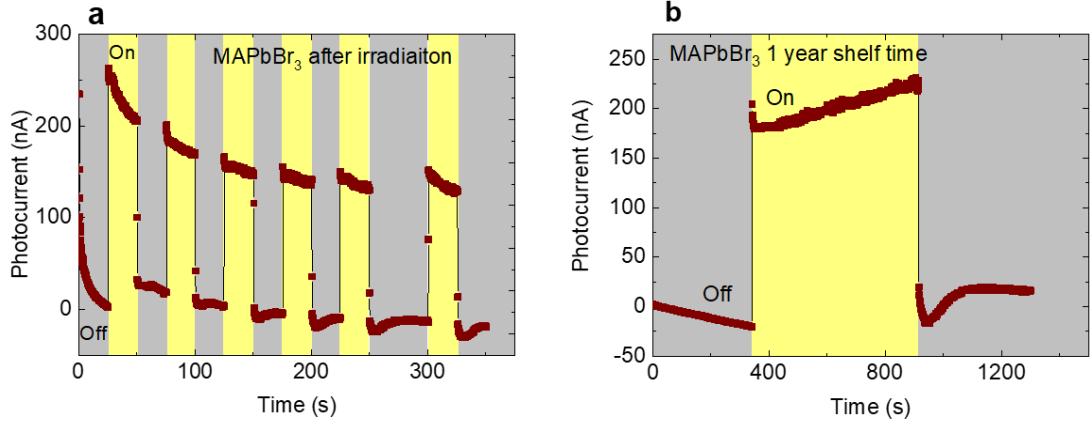

*Figure 7 – Electro-optical response of irradiated MAPbBr₃ device.* There is a clear photo-response right after absorbed 57 Sv H*(10) ambient total γ-dose (a). b) The photo-response of the device shows no degradation even after one-year shelf time.

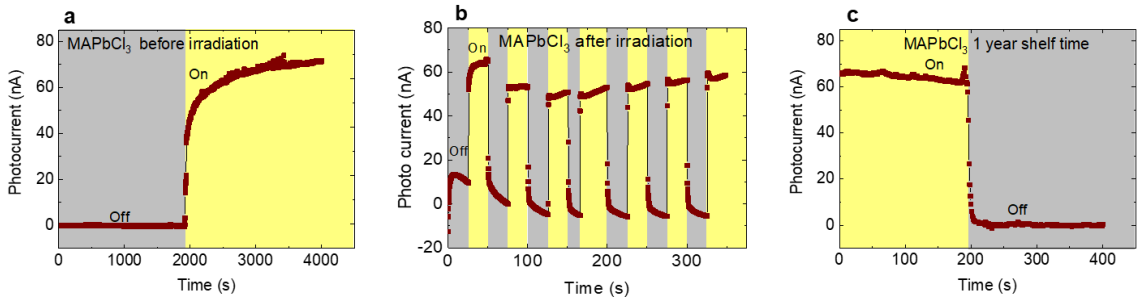

*Figure 8 – Electro-optical response of irradiated MAPbCl₃ device.* The photo-response before and right after absorbed 57 Sv H*(10) ambient total γ-dose is unchanged, showing radiation hardened behaviour of the MAPbCl₃ device.(a and b). c) The photo-response of the device shows no degradation even after one-year shelf time.

The current production efficiency is 1:0.2:0.015, for the I, Br, Cl analogues respectively. This also means that the MAPbBr₃ sample perform similar to the silicon based photovoltaics.[38] The gammavoltaic signal of both MAPbBr₃ and MAPbCl₃ was recorded in the same radiation conditions as for MAPbI₃ (Figure 9 and Figure 10).

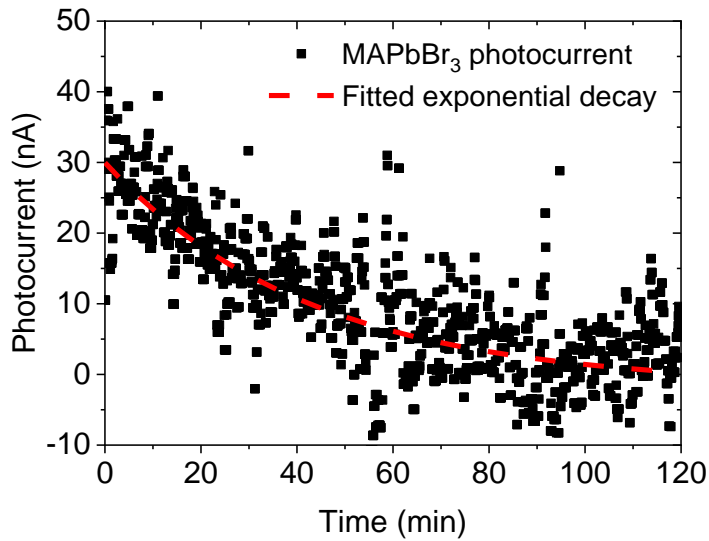

*Figure 9 – **Gammavoltaic effect of MAPbBr₃ exposed to the γ radiation of a shutdown nuclear reactor**. The γ-induced photocurrent follows the decreasing γ-dose rate with $\tau^{PC}$=2555±140 s time constant (extracted from the fit in red dashed line).*

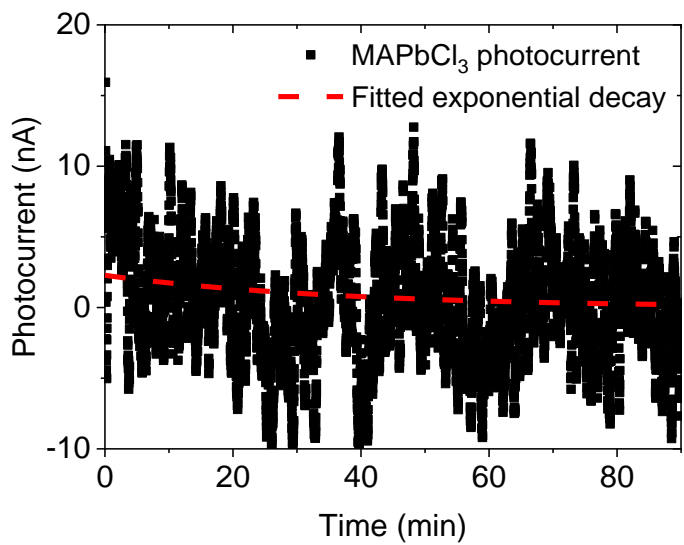

*Figure 10 – **Gammavoltaic effect of MAPbCl₃ exposed to the γ radiation of a shutdown nuclear reactor** The γ-induced photocurrent follows the decreasing γ-dose rate with $\tau^{PC}$=2220 s time constant, fit in red dashed line.*

At first sight, the radiation stability of HHPs might be surprising since these materials contain an organic component (methylammonium cations ($CH_3NH_3^+$)), which is supposed to be quite sensitive to high-energy irradiation. Indeed, several reports show the decomposition of MAPbI₃ to PbI₂ after even soft X-ray exposure[39]. On the other hand radiation tolerance[26] or even radiation enhanced behaviour is also frequently reported for HHPs.[27] This remarkable radiation resistance is referred to the "self-healing" i.e. the ability of the material to return to its original state after being damaged[27]. It is assumed that the vulnerability induced by the organic component is overcompensated by the ionic nature of the HHPs[40]. Also, the high defect-mobility[41] in HHPs might facilitate the self-healing of HHPs[42].

Finally, turn to the radiation shielding properties of MAPbI₃. Due to the high energy absorption coefficient, Pb-content and its relatively large density, it is natural to expect good radiation shielding performance. On a 0.034 cm³ crystal one cannot measure it precisely, but one can do reliable calculations with the MCNP code elaborated for similar purposes. Realistic MCNP calculations were performed on MAPbI₃, and various structural concretes. In these calculations, γ photons were sent through a shielding layer. The initial photon energy was varied in the 100 keV – 4 MeV range. The shielding performance was obtained from the photon spectrum on the other side of the material which was converted to a dose rate. By this MAPbI₃, ordinary concrete, and concrete with barite content were compared (Figure 3a). Additionally, calculations were run for the 10th value layer (a typical indicator for shielding efficiency), i.e. the average amount of material needed to absorb 90% of all radiation. The calculated values were compared with those of Si and GaAs, materials, which are commonly used as photodetectors, to check their shielding capabilities, as well (Figure 3b). All these calculations indicate MAPbI₃ as the most efficient shielding at most applied energies (Figure 3). This indicates the possibility of the multifunctional use of MAPbI₃ for example, as radiation shielding capable of backup energy production. For energy harvesting applications based on the 10th value layers of MAPbI₃ (Figure 3b), cm-scale layer thicknesses are needed which seems to be reachable by recent crystal growth methods[43].

## 4. Conclusions

In conclusion, a triple function was shown of MAPbI$_3$ in cases, where high densities of γ photons are present, e.g. nuclear fission, fusion, or outer space. The material can function as a detector and a current generator, as well as a shield for the outside environment. Radiation hardened γ-photon to free charge carrier conversion was observed in MAPbI$_3$ single crystal in a nuclear reactor. The crystals show no performance degradation due to irradiation up to 57 Sv H*(10) ambient γ-dose and 100 Sv/h γ-dose rate. Moreover, the observed gammavoltaic current did not show signs of saturation in the accessible dose rate range implying the possibility of dosimetry applications and direct energy harvesting capabilities of MAPbI$_3$ around nuclear power plants or radiation waste. Calculations revealed that the radiation shielding capabilities of MAPbI$_3$ are superior to concrete thus, it was proposedthat MAPbI$_3$ could form a basis of multifunctional radiation protection components where energy harvesting and environmental protection are simultaneously satisfied.

The potential relevant applications might include nuclear power plant failures, where an HHP-based gammavoltaic device could monitor the highly energetic, ionizing background by the generated photocurrent and simultaneously feed this current to the reactor cooling or safety system. The hours-long operational stability measured on symmetrically contacted single crystals is a good starting point for the realization of such devices in the future. An additional application could be nuclear batteries or high-altitude telecommunication balloon networks floating in the stratosphere like the Loon project[44] where lightweight energy source and radiation protection is required.

## 5. Funding

The work in Lausanne was partially supported by the Swiss National Science Foundation (Grant No. 513733) and ERC advanced grant "PICOPROP" (Grant No. 670918) and the ERC-POC grant PICOPROP4CT (Grant No. 790341).

Declarations of interest: none.

## 6. Appendix

*Table 1 – Overview of previous γ-irradiation experiments of HHPs.*

| Device structure | Sample size | Source | Activity | Dose rate (Sv/h) | Reference |
|---|---|---|---|---|---|
| Au / MAPbI$_3$ SC / PBCM:C$_{60}$ / Ga | 3.3 - 10 mm | Cesium-137<br>662 keV | 3.77 TBq | 29 | Dong et al.[15] |
| MAPbI$_3$ SC | 3 - 12 mm | Cesium-137<br>662 keV | 2.2 MBq | 0.07 | Yakunin et al.[45] |
| MAPbI$_3$ SC | 3 - 12 mm | Carbon-11<br>511 keV | 70 GBq | 1 | Yakunin et al.[45] |
| Cr / MAPbBr$_3$ SC / C$_{60}$ / BCP / Cr | 5×5×2 mm | Americium-241<br>59.6 keV | 29.6 kBq | 1.1E-4 | Xu et al.[46] |

| Material | Size | Source | Activity | | Reference |
|---|---|---|---|---|---|
| $Cs_xFA_{1-x}PbI_{3-y}Br_y$ (x=0–0.1, y=0–0.6) | 0.2 - 15 mm | Americium-241 59.6 keV | 0.4 MBq | - | Nazarenko et al.[47] |
| $Cs_xFA_{1-x}PbI_{3-y}Br_y$ (x=0–0.1, y=0–0.6) | 0.2 - 15 mm | Cesium-137 662 keV | 2.2 MBq | - | Nazarenko et al.[47] |
| Cr / $C_{60}$ / BCP / $MAPbBr_{2.94}Cl_{0.06}$ SC / Cr | 1.44×1.37×0.58 cm | Cesium-137 662 keV | 185 kBq | 1.4E-6 | Wei et al.[48,49] |
| $MAPbI_3$ | 3×3×3 mm | Nuclear reactor 30 keV – 3 MeV | 761 TBq | 100 | Present work |
| $MAPbBr_3$ | 3×3×3 mm | Nuclear reactor 30 keV – 3 MeV | 761 TBq | 100 | Present work |
| $MAPbCl_3$ | 3×3×3 mm | Nuclear reactor 30 keV – 3 MeV | 761 TBq | 100 | Present work |